\documentclass{edp-jp4}
\usepackage{graphicx}
\begin{document}

\title{Fate of the Wigner crystal on the square lattice } 
\author{ D.\ Baeriswyl}{\address{Department of Physics, University of
    Fribourg,  
Chemin du mus\'ee 3, CH-1700 Fribourg}
\author{S.\ Fratini}\address{LEPES--CNRS, 25 avenue des Martyrs,
    F-38042 Grenoble 
Cedex 9}%
\maketitle
\begin{abstract} The phase diagram of a system of electrons hopping 
on a square lattice and interacting through long-range Coulomb forces
is studied as a function of density and interaction strength. The
presence of a lattice strongly enhances the stability of the
Wigner crystal phase as compared to the case of the two-dimensional
electron gas.
\end{abstract}
%
\section{Introduction}
At very low densities the ground state of an
ensemble of charged particles is a Wigner crystal, with a triangular
structure if the motion is restricted to two dimensions. This 
striking phenomenon has
been observed for a very dilute electron system on the surface of 
liquid helium. In solids 
the already existing lattice periodicity interferes with the generally
different ordering tendency of the electrons.
We have previously studied this competition 
between Wigner crystallization and lattice commensurability for
spinless fermions hopping 
between the sites of a one-dimensional lattice and coupled pair-wise
by  long-range Coulomb forces \cite{valenzuela03,fratini04}. In this
case charge ordering is found to occur for any density and interaction 
strength, and only the amplitude of the modulation changes, from a Wigner 
superlattice for very large interaction strength to a small amplitude 
charge-density wave at weak coupling. 

Here we extend our considerations to the square 
lattice where a metal-insulator transition is expected to occur as 
a function of coupling strength. We first consider the classical limit
where the ground state structures show interesting variations as the
filling is varied. We then determine a crossover region
where the particles begin to spill over
to neighbouring sites due to quantum fluctuations. Finally we estimate
the critical interaction strength for which the Wigner crystal melts.
For small densities the continuum result is expected to remain valid,
while lattice commensurability effects become very important at larger 
densities.  

We consider the following Hamiltonian of spinless fermions,
\begin{equation}
H=-t\sum_{<i,j>}(c_i^{\dagger}c_j+c_j^{\dagger}c_i)+
\frac{1}{2}\sum_{i,j \ i\neq j}V_{i,j}(n_i-n)(n_j-n)\ ,
\label{hamiltonian}
\end{equation}
where $V_{i,j}=V/d_{i,j}$, $V$ being the nearest-neighbour interaction,
$n_i=c_i^{\dagger}c_i$
and $d_{i,j}$ is the distance between sites $i$ and $j$ of a square lattice. 
The effect of a 
constant compensating charge is explicitly included in the Coulomb term.
Our aim is to determine the ground state as a function of both $n$, the average
number of particles per site, and the relative interaction strength $V/t$.

\section{Classical limit} 

The classical problem ($t=0$) has been first addressed by Pokrovsky
and Uimin \cite{pokrovsky78}, who determined the ground-state structures
for a few rational fillings $n=1/q$, $q$ being a small integer.
Later Cocho proposed a simple recipe for generating the lowest-energy 
configurations for rational numbers $n=p/q, p>1$ \cite{cocho86}. In order 
to extract the effects of lattice commensurability
we have calculated numerically the energies for the simple rational
fillings $n=1/q$, assuming a Bravais lattice with
one electron per unit cell. 
The procedure can be directly generalised to other rational fillings. 

A perfect triangular structure on a host square lattice can
only be realized in the limit $n\to 0$. Thus
a measure of commensurability effects is the
difference between the energy of the ground state configuration and
that of such an ideal structure, given by
\begin{equation}
E_{ideal}=(-1.9605 \sqrt{n}+ 1.9501 n) V
\end{equation}
per particle. Our results are illustrated in Fig.\ 1 for the series $n=1/q$. 
The mismatch between the configuration on the lattice and the ideal structure 
is typically below $1\%$ at  fillings  $n<0.1$, but it clearly does not follow 
a monotonously increasing trend. Particularly stable ground state 
configurations appear for example at $n=1/8, 1/12, 1/15$, which are 
good approximants to the ideal triangular structure. Other configurations,
such as $n=1/7$, are less stable because of widely
different nearest-neigbour distances or bond angles.

\begin{figure}[htpb]
\centerline{\includegraphics[scale=0.7]{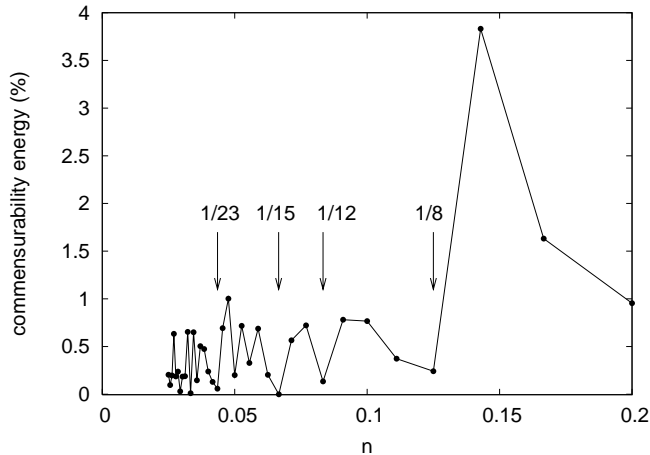}}
\caption{The deviation of the energy of the ground state configuration
  from that of a perfect triangular ordering, which gives a measure of the
  commensurability effects. Arrows mark the ground
  state structures which are good approximants to the optimal
  triangular structure. }
\end{figure}

\section{Quantum spreading}
A finite hopping term leads to excursions of particles away from their
classical equilibrium positions. 
For small values of $t$ the displacement will be mostly to nearest
neighbor sites, creating dipoles of length $a_0$, the lattice constant
of the host crystal. The energy of an individual dipole can be 
estimated in the continuum limit, i.e.\ for an ideal triangular structure.
We find
\begin{equation}
E_{dip}\approx 0.735n^{3/2}V.
\label{dip}
\end{equation}
To lowest order in perturbation theory the density on a nearest neighbor site 
of a superlattice point is given by 
\begin{equation}
\delta n = (t/E_{dip})^2.
\label{density}
\end{equation}
For appreciable spreading, say if the probability of finding the particle on
one of the four neighboring sites is about one half, we expect the discrete
nature of the underlying square lattice to become unimportant. Together
with Eqs.\ (\ref{dip}) and (\ref{density}) this gives a crossover region
around
\begin{equation}
V/t \approx 3.85n^{-3/2}
\label{crossover}
\end{equation}
between the parameter ranges of strong and weak discreteness effects.
\section{Quantum melting}
Our relation for the crossover region, Eq.\ (\ref{crossover}), shows that
for very small densities, discreteness effects are important only above
an extremely large interaction strength (or an extremely small hopping
parameter). Below this value the physics of our model is expected to be
equivalent to that of the ``two-dimensional electron gas'', for which
extensive numerical simulations have been performed. It is widely accepted
that this system undergoes a melting transition from a triangular Wigner
crystal to a correlated liquid at $r_s\approx 37$ \cite{tanatar89}. In our
model the corresponding parameter is obtained by identifying the tight-binding
spectrum for $k\to 0$ with a parabolic spectrum (with $\hbar^2/(2m^*a_0^2)=t$)
and by introducing an effective charge through the relation $(e^{*})^2/a_0=V$. 
Thus the effective Bohr radius is given by $a_B^*=2a_0t/V$. 
Together with the relation $\pi(r_sa_B^*)^2=a_0^2/n$ for the surface per
particle we obtain the 
following estimate for the melting transition at very small densities
\begin{equation}
V/t \approx 131 n^{1/2}.
\label{melting}
\end{equation}
Fig.\ 2 shows that for fillings above about 0.2 this line enters the region
where discreteness effects become important. For these fillings 
Eq.\ (\ref{melting}) can no longer be trusted and we have to go back to 
the original model,
Eq.\ (\ref{hamiltonian}). A simple way for estimating the melting line 
consists in using variational wave functions, one for the Wigner crystal, the
other for the metallic state. The most obvious choice is the ground state 
of the Coulomb term for the insulating phase 
(corresponding to the classical Wigner superlattice) compared to the ground 
state of the hopping term for the metallic phase (the restricted Hartree-Fock 
approximation). The energy of the classical (ideal) Wigner crystal is given 
by
\begin{equation}
E_{WC}=(-1.96n^{1/2}+1.95n)V,
\label{wc}
\end{equation}
while the Hartree-Fock solution can be expanded as
\begin{equation}
E_{HF}=(-1.51n^{1/2}+1.95n)V-(4-2\pi n)t+{\cal O}(n^{3/2}).
\label{hf}
\end{equation}
As expected, the former has lower energy for large $V/t$, while the
latter wins for small values of $V/t$. The energies are equal for
\begin{equation}
V/t\approx 8.9n^{-1/2}-14.0n^{1/2}.
\label{analytical}
\end{equation} 
This result (dotted line in Fig.\ 2) reproduces rather
well the numerical values (dots in the figure), which are based on the
actual energies of the different classical configurations and the (numerical) 
Hartree-Fock energies for the different densities.
\begin{figure}[htpb]
\centerline{\includegraphics[scale=0.7]{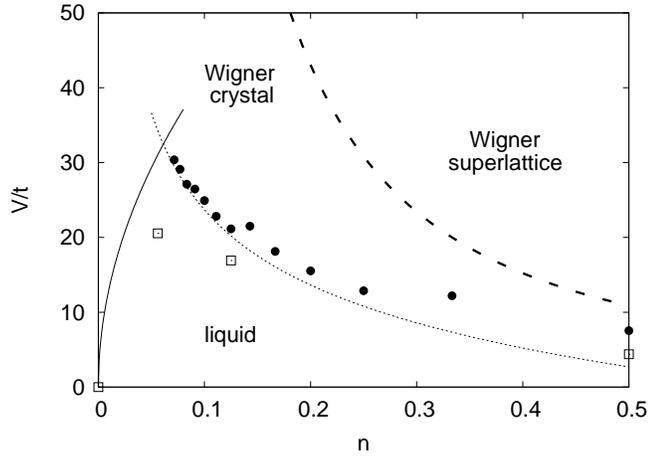}}
\caption{Phase diagram. The full line corresponds to the continuum limit
\cite{tanatar89}, the dotted line to Eq.\ (\ref{analytical}). The dots 
represent
our numerical results, the empty squares those of Ref.\ \cite{imada02}.}
\end{figure}

Our numerical results are expected to be quite reliable for densities
close to $n=0.5$ where they are not far away from the crossover line.
However, for lower densities we expect that correlation effects will 
stabilize the liquid phase
\cite{imada02}. Therefore we anticipate that below $n\approx 0.3$
the true melting line would be above our numerical results and then smoothly 
join that of the electron gas. At first sight this seems to contradict the
results of Noda and Imada \cite{imada02}, reproduced 
in Fig.\ 2 as empty squares, but one has to keep in mind that
by fine-tuning the underlying lattice they have stabilized artificially the 
Wigner crystal phase. 

\section{Discussion}
We have studied the stability of the 
Wigner crystal phase on a square lattice. The phase diagram (interaction
strength $V/t$ versus density $n$) shows different regimes. For very small 
densities ($n<0.1$) the melting line is that of the 
two-dimensional electron gas. Strong deviations occur close to $n=0.5$ where 
the Wigner crystal is stabilized by the lattice (the motion is more and 
more hindered as the density increases). In-between, the melting
line is expected to go through a maximum. Despite the stabilizing effects
of the lattice the Wigner crystal is still only stable for very large 
interaction strengths ($V_{c,min}\approx 6t$). Therefore Coulomb interactions
alone are unlikely to promote charge ordering in layered cuprates where 
$V\approx t$. In these and related materials other effects, in particular 
three-dimensionality \cite{rastelli05}, electron-lattice coupling,  
counter-ions or magnetic exchange may be necessary to stabilize charge-ordered
structures.



\end{document}